\documentclass[12pt, a4, reqno]{amsart}
\usepackage{natbib}
\usepackage[colorlinks=true, urlcolor=blue, linkcolor=blue, citecolor=blue]{hyperref}
\usepackage{amsmath,amssymb,centernot,mathtools}
\usepackage{amsthm}
\usepackage{graphicx}
\usepackage[lofdepth,lotdepth]{subfig}
\usepackage{caption}
\usepackage{enumitem}
\usepackage{color}
\usepackage[table]{xcolor}
\usepackage[calc]{picture}
\usepackage[english]{babel}
\usepackage[margin=1in]{geometry}
\usepackage{secdot}
\usepackage{float}
\usepackage{tikz}
\usepackage{bm}
\usepackage[UKenglish]{isodate}

\usetikzlibrary{arrows.meta,shapes}
\usetikzlibrary{arrows,shapes.arrows,shapes.geometric,shapes.multipart,
decorations.pathmorphing,positioning,shapes.swigs}

\usepackage{setspace}
\usepackage{tcolorbox}
\usepackage[outline]{contour}
\usepackage[markers,nolists]{endfloat}
\usepackage{algcompatible}
\usepackage{algorithm,algpseudocode,float}
\usepackage{pdflscape}
\usepackage{cancel}
\usepackage{xr}
\usepackage{etoolbox}

\theoremstyle{definition}
\newtheorem{assumption}{Assumption}

\newtheorem{theorem}{Theorem}
\newtheorem{proposition}{Proposition}
\newtheorem{remark}{Remark}

\newcommand\independent{\protect\mathpalette{\protect\independenT}{\perp}}
\def\independenT#1#2{\mathrel{\rlap{$#1#2$}\mkern2mu{#1#2}}}

\newcommand\norm[1]{\left\lVert#1\right\rVert}

\newcommand{\indep}{\perp\mkern-9.5mu\perp}

\makeatletter
\newenvironment{breakablealgorithm}
  {
   \begin{center}
     \refstepcounter{algorithm}
     \hrule height.8pt depth0pt \kern2pt
     \renewcommand{\caption}[2][\relax]{
       {\raggedright\textbf{\ALG@name~\thealgorithm} ##2\par}%
       \ifx\relax##1\relax 
         \addcontentsline{loa}{algorithm}{\protect\numberline{\thealgorithm}##2}%
       \else 
         \addcontentsline{loa}{algorithm}{\protect\numberline{\thealgorithm}##1}%
       \fi
       \kern2pt\hrule\kern2pt
     }
  }{
     \kern2pt\hrule\relax
   \end{center}
  }
\makeatother

\makeatletter
\newcommand*{\addFileDependency}[1]{
  \typeout{(#1)}
  \@addtofilelist{#1}
  \IfFileExists{#1}{}{\typeout{No file #1.}}
}
\makeatother

\newcommand*{\myexternaldocument}[1]{%
    \externaldocument{#1}%
    \addFileDependency{#1.tex}%
    \addFileDependency{#1.aux}%
}

\myexternaldocument{webappendix}

\doublespace

\begin{document}

\title[Causal Inference with Incomplete Exposure and Confounders]{Estimating Average Causal Effects with Incomplete Exposure and Confounders}

\author[Wen \& McGee]{Lan Wen$^\ast$ and Glen McGee\\ 
\textit{Department of Statistics and Actuarial Science, University of Waterloo, Waterloo, Ontario, Canada} \\ 
{lan.wen@uwaterloo.ca} \\}

\maketitle
\begin{center}\textit{First draft}: \printdate{29.10.2023}  ~~~ \textit{Current draft}: \printdate{23.06.2025}\end{center}


\begin{abstract}
{Standard methods for estimating average causal effects require complete observations of the exposure and confounders.  In observational studies, however, missing data are ubiquitous. Motivated by a study on the effect of prescription opioids on mortality, we propose methods for estimating average causal effects when exposures and potential confounders may be missing. We consider missingness at random and additionally propose several specific missing not at random (MNAR) assumptions. Under our proposed MNAR assumptions, we show that the average causal effects are identified from the observed data and derive corresponding {influence functions in a nonparametric model}, which form the basis of our proposed estimators. Our simulations show that standard multiple imputation techniques paired with a complete data estimator is unbiased when data are missing at random (MAR) but can be biased otherwise. For each of the MNAR assumptions, we instead propose doubly robust targeted maximum likelihood estimators (TMLE), allowing misspecification of either (i) the outcome models or (ii) the exposure and missingness models. The proposed methods are suitable for any outcome types, and we apply them to a motivating study that examines the effect of prescription opioid usage on all-cause mortality using data from the National Health and Nutrition Examination Survey (NHANES).}
\end{abstract}\noindent {causal inference; double robustness; missing not at random; multiple imputation; outcome-independent missingness; targeted learning.}

\sloppy
\clearpage
\section{Introduction}\label{s:intro}
Observational studies are often used to investigate causal questions in epidemiology. When outcomes, exposures and confounders are all completely observed, estimating causal effects proceeds via standard estimation techniques like parametric g-formula \citep{Taubman2009,Lajous2013,Lodi2015}, propensity score-based methods \citep{Hernan2000,Cain2010, Neugebauer2014}, or doubly robust methods \citep{Bang2005,Tran2019,Wen2023}. A limitation of observational studies, however, is that data are often missing. Consider, for example, a motivating analysis of the 1999---2004 cycle of the National Health and Nutrition Examination Survey (NHANES), which found that opioid prescriptions increased risk of all-cause mortality  \citep{inoue2022causal}. 
Our interest lies in estimating the average causal effect of opioid prescription on 5-year all-cause mortality among the elderly, and the data include several potential confounders. 
The NHANES data is linked to the National Death Index mortality data, so while the outcome of interest is fully observed, missing data arise in the exposure and potential confounders.
More specifically, 24\% of observations were missing either the exposure of interest or a confounder in the analytic sample, so standard methods that rely on data being complete may not apply.

Extensions have been proposed to accommodate either missing exposures \citep{williamson2012doubly,zhang2016causal,kennedy2020efficient,sperrin2020multiple} or missing confounders \citep{qu2009propensity,crowe2010comparison,mitra2011estimating,seaman2014inverse,williamson2012doubly,evans2020coherent,levis2022robust,ross2022reflection,d2000estimating,ding2014identifiability,yang2019causal,guan2019unified,sun2021semiparametric,miao2018identification,blake2020estimating,blake2020propensity,lu2018propensity,mattei2009estimating}. When the exposure is missing at random (MAR; \citealp{rubin1976inference,Seaman2013,mohan2021graphical}) and confounders are fully observed, several authors proposed augmented inverse probability weighted estimators \citep{williamson2012doubly,zhang2016causal,kennedy2020efficient}. 
When confounders are MAR and exposure is fully observed, existing methods include multiple imputation (MI) \citep{qu2009propensity,crowe2010comparison,mitra2011estimating,seaman2014inverse}, influence function-based estimators \citep{scharfstein1999adjusting,evans2020coherent,levis2022robust} and inverse probability weighting (IPW; \citealp{ross2022reflection}). 
  When confounders are missing not at random (MNAR), further assumptions are needed. Under an outcome-independence assumption \citep{ding2014identifiability}, proposed approaches include two-stage least squares estimation \citep{yang2019causal}, imputation strategies \citep{guan2019unified}, and doubly robust estimation \citep{sun2021semiparametric}---though all rely on a rank constraint (also referred to as a completeness condition) that precludes binary outcomes. Other authors assumed existence of a shadow variable \citep{miao2018identification} or that covariates act as confounders only when they are observed \citep{rosenbaum1984reducing,d2000estimating,blake2020propensity,blake2020estimating}.

Missingness is not restricted to only exposures or only confounders, yet there exists limited guidance on estimating average causal effects when both exposure and confounders may be missing.  
Several authors provided algorithms for identifiability and/or testability based on graphical models, but did not discuss plausibility of the graphs nor estimation in causal settings \citep{mohan2013graphical,mohan2014testability,bhattacharya2020identification,nabi2023testability}. Starting from  Directed Acyclic Graphs (DAGs) depicting missingness mechanisms (m-DAGs), \cite{moreno2018canonical} discussed identifiability of causal quantities but limited discussion of estimation to available-case analysis and MI (see \citealp{dashti2021handling}). In particular, they noted that efficient estimation ``require tailoring for each m-DAG and recoverable parameter and have not commonly been used''.

The focus of this paper is on estimating average causal effects when both exposure and potential confounders may be partially observed. 
We begin with one of several assumptions about the missingness mechanism(s): besides a MAR assumption, we propose several variations of the outcome-independence assumption. Motivated by the NHANES analysis, the proposed MNAR assumptions permit missingness in the exposure---opioid prescription---to depend on missing potential confounders and the missing exposure itself, and allows shared unmeasured common causes between these aforementioned variables. Importantly, none of the assumptions restrict the types of outcomes allowed. {Under each set of assumptions, we derive identifying formulae, and corresponding influence functions under a nonparametric model}. We show that standard MI (e.g., Joint Multivariate Normal MI, Fully Conditional Specification MI) yields valid estimates when data are MAR but can be biased otherwise. For each of the MNAR assumptions, we propose targeted maximum likelihood estimators (TMLE) that are doubly robust, allowing misspecification of either (i) the outcome models or (ii) the exposure and missingness models.

\section{Setup: Causal estimand and  assumptions}\label{s:standard}

{Suppose we collect $n$ fully observed, independent and identically distributed observations. Under this setting,} let $Y$ denote an outcome of interest, $A$ an exposure variable taking values in $\mathbb{N}$, and $L$ a vector of potential confounders. For ease of exposition, we assume throughout that the outcome $Y$ is binary and the covariates $L$ are discrete, but methods described herein are generalizable to other outcomes types and to continuous covariates.
Let $Y^a$ denote the counterfactual outcome variable if, possibly contrary to fact, the exposure had taken a value $a$ for $a\in \textrm{supp}(A)$, and {let $E(Y^{a})$ denote the corresponding average counterfactual outcome of interest, had exposure taken a value $a$ for all individuals in a population}. 
{To quantify average causal effects, we consider contrasts of mean counterfactual outcomes: e.g., $E(Y^{a=1})-E(Y^{a=0})$.} 

{Throughout we will assume that statistical independencies in the data are faithful to the DAG under study \citep{verma2022equivalence}.
We also assume no interference or the stable unit-treatment value assumption (SUTVA; \citealp{rubin1980randomization}), in addition to the following standard causal assumptions stated below:} 
\begin{assumption}[Consistency] If $A=a$, then
\label{ass:consG}
$
Y^{a} = Y, \quad 
$
\end{assumption}
\begin{assumption}[{Conditional} Exchangeability] 
\label{ass:NUC}
$
Y^a\indep A\mid L.
$
\end{assumption}
{When the exposure or confounders are partially missing, we must make additional assumptions in order to identify $E(Y^{a})$, for any $a\in \textrm{supp}(A)$.}
In this paper, we allow the exposure $A$ and confounders $L$ to be partially missing; in particular we let $L_M$ represent the subvector of confounders that are partially missing, and $L_O$ represent the subvector of confounders that are fully observed for all individuals; i.e., $L=(L_M,L_O)$. In Section \ref{s:simultaneous} and \ref{sec:separate}, we {consider various missingness mechanisms for $A$ and $L_M$}. In Section \ref{s:sims}, we show simulation results and illustrate theoretical findings empirically, and in \ref{s:analysis}, we apply the new methods to study the effect of opioid prescriptions on mortality using data from NHANES.
We conclude with a discussion in Section \ref{s:discussion}.

\section{Data that are missing at random}\label{s:simultaneous}
{Due to its convenient implementation, standard off-the-shelf MI is frequently used in social and epidemiological studies to handle missing data in estimating average causal effects \citep{lopoo2005incarceration,leslie2014cervical, lippold2014investigating,kreif2016evaluating,decruyenaere2020obesity,anthony2021examining,dashti2021handling,inoue2022causal}. This, however, implicitly requires that data are MAR. In this section, we briefly discuss identification and estimation under this assumption.}

{Separating the observation indicators for $L_M$ and $A$ would induce non-monotone missing data, for which an MAR mechanism is challenging to justify scientifically \citep{Robins1997, Vansteelandt2007, sun2018inverse}. Hence, in this section, we treat $A$ and $L_M$ as either entirely missing or fully observed for each subject. To account for this, let $R$ denote the observation status of $(L_M,A)$ such that $R=1$ if both $L_M$ and $A$ are observed and $R=0$ otherwise.} As such, the observed data for an individual are $O = (L_O,R, R L_M, R A, Y)\sim P$. {We consider an (everywhere) MAR assumption (\citealp{Seaman2013}; see also a similar v-MAR assumption in \citealp{mohan2021graphical}) for both exposure and confounders as well as a positivity assumption for this missingness process:}
\begin{assumption}[MAR]
\label{ass:MARwithboth}
    $R \indep \{A,L_M\}|Y,L_O.$
\end{assumption}
\begin{assumption}[Positivity: MAR]
\label{ass:posMAR}
      $P(R=1\mid L_O=l,Y=y)>0,~~\forall (l_O,~y)\in\textup{supp}({L_O},~{Y}).$
\end{assumption}
Under assumptions \ref{ass:MARwithboth} and \ref{ass:posMAR}, the {average counterfactual outcome $E(Y^{a})$} is identified by the following:
\begin{align}
   \Psi_{MAR}^{a} 
    & = E\left[\frac{\sum_y yp(L_M,a\mid L_O, y, R=1)p(y\mid L_O)}{\sum_{y} p(L_M,a\mid L_O, y, R=1)p(y\mid L_O)}   \right] 
 \label{eq:id1}
 \\& = E\left[\frac{\beta(L)}{\gamma(L)}\frac{R}{P(R=1\mid L_O,Y)}\right],\nonumber
\end{align}
where $\beta(L) = {\sum_y yp(L_M,a\mid L_O, y, R=1)p(y\mid L_O)}$ and $\gamma(L) = \sum_{y} p(L_M,a\mid L_O, y, R=1)p(y\mid L_O)$. This generalizes the results of \citet{levis2022robust}, which considered missing confounders only.

\begin{figure}[htbp!]
	\centering
  \subfloat[]{\begin{minipage}{0.45\textwidth}
	\centering
	\begin{tikzpicture}
		\node (LM) at (-3,1.5) {$L_M$};
            \node (LO) at (-3,-1.5) {$L_O$};
		\node (Y) at (2,0) {$Y$};
		\node (A) at (-2,0) {$A$};
		\node (R) at (0,0) {$R$};
		\draw [->] (LM) edge (A);
		\draw [->] (LM) edge[bend left] (Y);
                \draw [<->] (LM) edge[bend left=45, dashed] (Y);
                \draw [<->] (LM) edge[bend right, dashed] (LO);
            \draw [->] (LO) edge (A);
		\draw [->] (LO) edge[bend right] (R);
		\draw [->] (LO) edge[bend right] (Y);
                \draw [<->] (LO) edge[bend right=45, dashed] (Y);
  		\draw [->] (A)  edge[bend left] (Y);
	\end{tikzpicture}
  \end{minipage} }
 \subfloat[]{\begin{minipage}{0.45\textwidth}
 \centering
 	\begin{tikzpicture}
		\node (LM) at (-3,1.5) {$L_M$};
            \node (LO) at (-3,-1.5) {$L_O$};
		\node (Y) at (0,0) {$Y$};
		\node (A) at (-2,0) {$A$};
		\node (R) at (2,0) {$R$};
		\draw [->] (LM) edge (A);
		\draw [->] (LM) edge[bend left] (Y);
                \draw [<->] (LM) edge[bend right, dashed] (A);
            \draw [->] (LO) edge (A);
		\draw [->] (LO) edge[bend right] (R);
		\draw [->] (LO) edge[bend right] (Y);
                \draw [<->] (LO) edge[bend right=45, dashed] (R);
  		\draw [->] (A)  edge (Y);
     \draw [->] (Y)  edge (R);
	\end{tikzpicture}
 \end{minipage} }
	\caption[.]{DAGs satisfying assumption \ref{ass:MARwithboth} (MAR). Dashed lines represent potential unmeasured common causes.  }
\label{fig:MARwithboth1}
\end{figure}
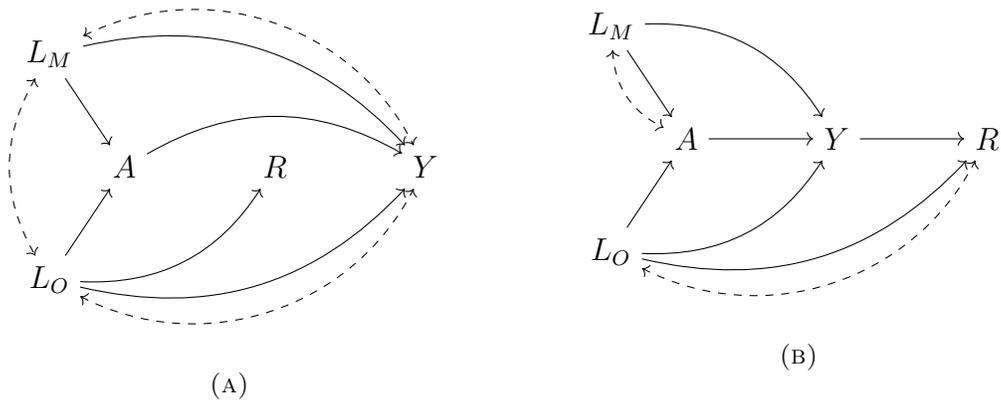

Figure \ref{fig:MARwithboth1} displays examples of causal DAGs that satisfy assumption \ref{ass:MARwithboth} (note that these are non-exhaustive; other DAGs are possible under our assumptions). The MAR assumption allows missingness status $R$ to depend on observed confounders $L_O$. 
{For instance, in our data application, `income' and `marital status' are partially observed. It is possible that individuals in the NHANES study had privacy concerns regarding these two variables, and privacy concerns, `income' and `marital status' are likely affected by one's age, which is fully observed.}
Interestingly, the temporal ordering of $R$ and $Y$ has important implications when the MAR assumption \ref{ass:MARwithboth} and conditional exchangeability assumption \ref{ass:NUC} hold. In our DAG examples, the outcome may affect missingness in Figure \ref{fig:MARwithboth1}[b] (e.g., in a retrospective study, one's health outcome may affect how likely they are to stay in a study and report a past exposure status), but missingness may not affect the outcome in Figure \ref{fig:MARwithboth1}[a]. 
{The latter is a property that we assume throughout (see \cite{srinivasan2023graphical} who relax this assumption).}
In general, the MAR assumption \ref{ass:MARwithboth} can be  restrictive, because neither the missing exposure $A$ nor missing confounders $L_M$ are allowed to affect $R$ . 

{In Web Appendix E, we derive an influence function for $\Psi_{MAR}^{a}$ given by identifying formula \eqref{eq:id1} under a nonparametric model for $P$. In contrast to the other identifying formulae we present, the form of the influence function requires specifying the joint distribution of $L_M$ and $A$, hence using this as a basis for estimation is practically challenging. Alternatively, we can leverage the MAR assumption and exploit standard MI with available software -- a common practice in social and epidemiological studies --  then apply existing TMLE approach to the imputed complete data (\citealp{van2011targeted}; see Web Appendix F for more details).} 
{While MI is easy to implement, the MAR assumption can be overly restrictive, as described above. In the next section, we propose assumptions that deviate from MAR and hence correspond to MNAR in the missing data literature \citep{LittleRubin2002}.}

\section{Data that are not missing at random}
\label{sec:separate}

We now consider separating the observation indicators of exposure $A$ and confounders $L_M$.
In the following, we require that the missingness mechanism be \textit{outcome-independent}: {$R \indep Y|A,L$} \citep{Zhao2015,miao2018identification, yang2019causal}, but do not assume any rank conditions.
{The outcome-independence assumption may be reasonable e.g., in prospective studies where covariates and exposure are measured before the outcome occurs \citep{yang2019causal}, but may be less plausible if the reasons for missingness (e.g., poor underlying health conditions) also affect the outcome of interest.}

{In what follows, let $R_A$ denote the observation indicator for exposure ($R_A=1$ when $A$ is observed and $R_A=0$ otherwise). 
Throughout this section, we assume an \textit{$R_A$-outcome independence condition}, namely that {$R_A \indep Y|A,L$}.
In addition, we posit two classes of missing-data mechanisms for the partially observed covariates $L_M$ that fall outside the MAR framework (i.e., are MNAR).
These two missingness mechanisms impose additional, stronger assumptions for the missingness in $L_M$ than those imposed for the missingness in the exposure.}

\subsection{Identification and estimation under an MNAR mechanism}
\label{sec:MNARA}
 Let $R_L$ denote the observation indicator for covariates $L_M$ ($R_L=1$ when $L_M$ is entirely observed and $R_L=0$ otherwise). In this case, $R_A$ may differ from $R_L$, and we now observe $O = (L_O,R_A,R_L, R_L L_M, R_A A, Y)\sim P$. Consider the following assumptions:
\begin{assumption}[MNAR-A]
\label{ass:MNAR-A} ~
\begin{enumerate}
    \item {Exposure missingness: $R_A \indep Y|A,L$}
    \item {Covariate missingness: $R_L \indep Y|A,L,R_A$ and $R_L \indep L_M|L_O.$}
\end{enumerate}
\end{assumption}
\begin{assumption}[Positivity: MNAR-A]
\label{ass:posMNAR-A}
         $$P(A=a\mid R=1, L=l)>0,~~\forall l\in\textup{supp}({L}), ~ \text{and} ~ 
    P(R_L=1\mid L_O=l_O)>0,~~\forall l_O\in\textup{supp}({L_O}).$$
\end{assumption}

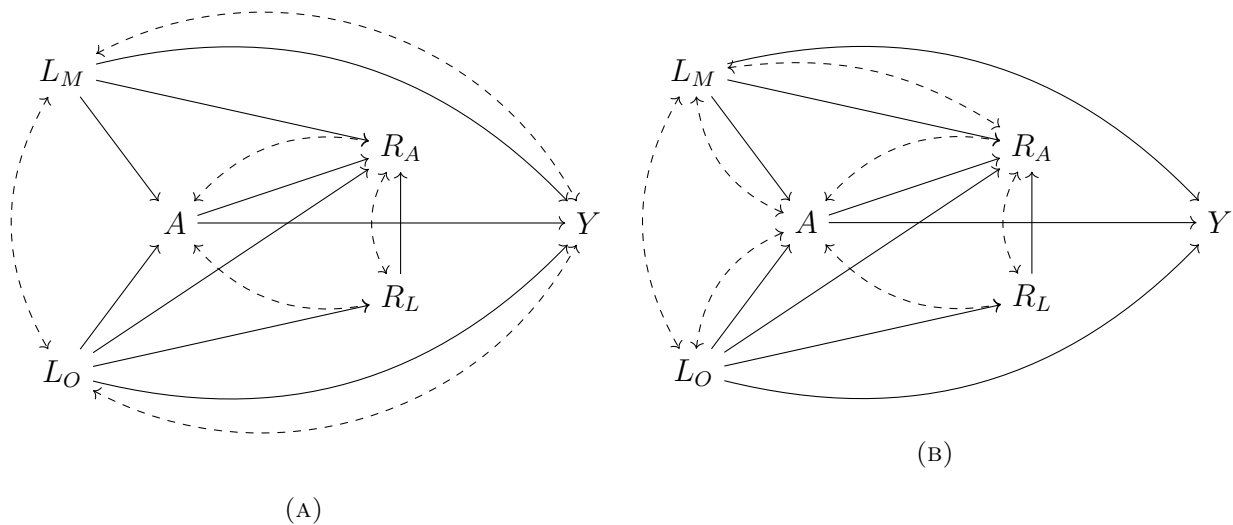
\begin{figure}[!htbp]
	\centering
  \subfloat[]{\begin{minipage}{0.5\textwidth}
   \centering
	\scalebox{1}{
	\begin{tikzpicture}
		\node (LM) at (-3,2) {$L_M$};
            \node (LO) at (-3,-2) {$L_O$};
		\node (Y) at (4,0) {$Y$};
		\node (A) at (-1.5,0) {$A$};
		\node (RA) at (1.5,1) {$R_A$};
        \node (RL) at (1.5,-1) {$R_L$};
		\draw [->] (LM) edge (A);
		\draw [->] (LM) edge[bend left] (Y);
                \draw [<->] (LM) edge[bend left=45, dashed] (Y);
                \draw [<->] (LM) edge[bend right, dashed] (LO);
            \draw [->] (LO) edge (A);
		\draw [->] (LO) edge (RA);
		\draw [->] (LO) edge[bend right] (Y);
                \draw [<->] (LO) edge[bend right=45, dashed] (Y);
  		\draw [->] (A)  edge (Y);
		\draw [->] (A) edge (RA);
        \draw [->] (LM) edge (RA);
        \draw [->] (RL) edge (RA);
        \draw [->] (LO) edge (RL);
        \draw [<->] (A) edge[bend left, dashed] (RA);
         \draw [<->] (A) edge[bend right, dashed] (RL);
        \draw [<->] (RL) edge[bend left, dashed] (RA);
	\end{tikzpicture} }
  \end{minipage} }
      \subfloat[]{\begin{minipage}{0.5\textwidth}
   \centering
  	\scalebox{1}{
	\begin{tikzpicture}
		\node (LM) at (-3,2) {$L_M$};
            \node (LO) at (-3,-2) {$L_O$};
		\node (Y) at (4,0) {$Y$};
		\node (A) at (-1.5,0) {$A$};
		\node (RA) at (1.5,1) {$R_A$};
        \node (RL) at (1.5,-1) {$R_L$};
		\draw [->] (LM) edge (A);
		\draw [->] (LM) edge[bend left] (Y);
                \draw [<->] (LM) edge[bend right, dashed] (LO);
            \draw [->] (LO) edge (A);
		\draw [->] (LO) edge (RA);
		\draw [->] (LO) edge[bend right] (Y);
  		\draw [->] (A)  edge (Y);
		\draw [->] (A) edge (RA);
        \draw [->] (LM) edge (RA);
        \draw [->] (RL) edge (RA);
        \draw [->] (LO) edge (RL);
        \draw [<->] (A) edge[bend left, dashed] (RA);
         \draw [<->] (A) edge[bend right, dashed] (RL);
         \draw [<->] (LM) edge[bend left=20, dashed] (RA);
        \draw [<->] (RL) edge[bend left, dashed] (RA);
        \draw [<->] (LM) edge[bend right, dashed] (A);
        \draw [<->] (LO) edge[bend left, dashed] (A);
	\end{tikzpicture} }
  \end{minipage} } 
	\caption[.]{Example  DAG satisfying MNAR-A assumption \ref{ass:MNAR-A}. Dashed lines represent potential unmeasured common causes.}
  \label{fig:assMNAR-A3}
\end{figure}

Under assumptions \ref{ass:MNAR-A} and \ref{ass:posMNAR-A}, $E(Y^{a})$ can be identified using the following functional:
\begin{align}
   \Psi_{MNAR-A}^{a} =  \sum_l E(Y\mid A=a, R_A=1,R_L=1, L)p(l_M\mid l_O, R_L=1)p(l_O). \label{eq:idMNAR-A}
\end{align}
Figure \ref{fig:assMNAR-A3} shows two DAGs satisfying these assumptions, where $R_A$ can be affected by $A$ and $L_M$, and both $R_A$ and $R_L$ can share common causes with $A$. This might be the case in observational studies if exposure information is sensitive, such as opioid intake.
For instance, it is possible that opioid usage affects one's perceived social status, which in turn can affect whether someone reports their opioid usage. It is also possible that an individual's perception of certain social factors (e.g., societal and cultural norms) can affect whether they take opioids (exposure) and whether they report their exposure and covariate statuses.

\subsubsection{Estimation}
{We first derive the corresponding influence function for the functional $\Psi_{MNAR-A}^{a}$ given by identifying formula (\ref{eq:idMNAR-A}) under a nonparametric model $\mathcal{M}_{np}$ for $P$. The resulting influence function is given by}:
\begin{align}
    \phi^1_{P_{\bf{MNAR-A}}}(O) = \frac{I(A=a,R_A=1,R_L=1)}{\pi_A(L)\pi_{R_A}(L) \pi_{R_L}(L_O)}\{Y-T_1(L)\} + \frac{I(R_L=1)}{\pi_{R_L}(L_O)}\{T_1(L)-{T}_0(L_O)\} + {T}_0(L_O) - \Psi_{MNAR-A}^{a}
    \label{eq:eifmnar12}
\end{align} 
where  $\pi_A(L) = P(A=a\mid R=1, L)$, 
$\pi_{R_A}(L)=P(R_A=1\mid L,R_L=1)$, $\pi_{R_L}(L_O)=P(R_L=1\mid L_O)$, 
$T_1(L) = E(Y\mid A=a, R=1,L)$, and ${T}_0(L_O) = E(T_1(L)\mid L_O,R_L=1)$.  
With this influence function in hand, we can construct consistent estimators of $\Psi_{MNAR-A}^{a}$, such as the solution to estimating equations or the TMLE. We propose a TMLE via Algorithm in Table \ref{tab:tmle_MNAR-A}, which conveniently reduces to the complete-data TMLE when there is no missing data \citep{van2011targeted}.

\begin{table}
\caption{\label{tab:tmle_MNAR-A}}
\begin{breakablealgorithm}
\renewcommand{\theenumi}{\Alph{enumi}}   
\caption{{TMLE-A under MNAR-A}}         
\begin{algorithmic} [1]     
\setstretch{1.5}
\item Obtain estimates $\hat\pi_A(L)$, $\hat\pi_{R_A}(L)$ and $\hat\pi_{R_L}(L_O)$ of  $\pi_A(L)$, $\pi_{R_A}(L)$ and $\pi_{R_L}(L_O)$, respectively. 
\item \textit{Obtain initial prediction of $T_1(L)$, $\hat T_1^0(L)$}: 

\noindent Among those with $R=1$, fit a regression model $\eta_1(A,L;\kappa_1)=g^{-1}([A,L]'\kappa_1)$ by regressing $Y$ on $A$ and $L$. Alternatively, we can regress $Y$ on $A$ and $L$ in those whose $R=1$ without having to stratify on $A$. Obtain predictions $\hat T_1^0(L) = \eta_1(A,L;\hat\kappa_1)$ for these individuals.
\item \textit{Targeting step for $T_1(L)$ to obtain updated prediction $\hat T_1^\ast(L)$}: 

\noindent Among those with $(A,R)=(a,1)$, regress $Y$ on an intercept with observational weight $\{\hat\pi_A(L)\hat\pi_{R_A}(L)\hat\pi_{R_L}(L_O)\}^{-1}$ and an offset given by $g\{\hat T_1^0(L)\}$, i.e., solve for $\epsilon_1$ in 
\begin{equation*}
   \mathbb{P}_n \left\{ \frac{I(A=a,R_A=1,R_L=1)}{\hat{\pi}_A(L)\hat\pi_{R_A}(L)\hat\pi_{R_L}(L_O)}\left(Y - g^{-1}\left[g\{\hat T_1^0(L)\}+\epsilon_1\right]\right)\right\} = 0
\end{equation*}
 Among those with $R_L=1$, predict $T_1(L)$ using $\hat T_1^\ast(L) = g^{-1}\left[g\{\hat T_1^0(L)\}+\hat\epsilon_1\right]$.
\item \textit{Obtain initial prediction of ${T}_0(L_O)$, $\hat T_0^0(L_O)$}: 

\noindent Among those with $R_L=1$, fit a regression model $\eta_0(L_O;\kappa_0)=g^{-1}(L_O'\kappa_0)$ by regressing $\hat T_1^\ast(L)$ on $L_O$. 
Obtain predictions $\hat T_0^0(L_O) = \eta_0(L_O;\hat\kappa_0)$  for these individuals.
\item \textit{Targeting step for ${T}_0(L_O)$} to obtain updated prediction $\hat T_0^\ast(L_O)$: 

\noindent Among those with $R_L=1$, regress $\hat T_1^\ast(L)$ on an intercept with observational weight $\hat\pi_{R_L}(L_O)^{-1}$ and an offset given by $g\{\hat T_0^0(L_O)\}$, i.e., solve for $\epsilon_0$ in 
\begin{equation*}
   \mathbb{P}_n \left\{ \frac{I(R_L=1)}{\hat\pi_{R_L}(L_O)}\left(\hat T_1^\ast(L) - g^{-1}\left[g\{\hat T_0^0(L_O)\}+\epsilon_0\right]\right)\right\} = 0
\end{equation*}
Predict ${T}_0(L_O)$ using $\hat{ {T}}_0^\ast(L_O)=g^{-1}\left[g\{\hat T_0^0(L_O)\}+\hat\epsilon_0\right]$ for all observations.
\item Calculate the TMLE-A  estimator $\widehat{\Psi}_{\bf{TMLE,MNAR-A}}^a=\mathbb{P}_n\{\hat{{T}}_0^\ast(L_O)\}$.
\end{algorithmic}
\end{breakablealgorithm}
\vphantom{}	
\end{table}

{Throughout, we will use $\mathcal{M}$ with subscripts ``${T_1}$'' and ``${T_0}$'' to denote models with the correct specification of the nuisance functions defined by $T_1(L)$ and $T_0(L_O)$, and $\mathcal{M}$ with subscripts 
``${\pi_{A,R_A}}$'' and ``${\pi_{R_L}}$'' to denote models with the correct specification of $P(A=a, R_A=1\mid R_L=1, L)$ and $P(R_L=1\mid L_O)$, respectively.}
  \begin{proposition}
      {The proposed TMLE-A is doubly robust in the sense that it is consistent for $\Psi_{MNAR-A}^{a}$ under ($\mathcal{M}_{T_1}\cap \mathcal{M}_{T_0})\cup (\mathcal{M}_{\pi_{A,R_A}}\cap \mathcal{M}_{\pi_{R_L}})$. Moreover, the TMLE-A estimator has an influence function given by \eqref{eq:eifmnar12} under ($\mathcal{M}_{T_1}\cap \mathcal{M}_{T_0})\cap (\mathcal{M}_{\pi_{A,R_A}}\cap \mathcal{M}_{\pi_{R_L}})$, and thus it achieves the nonparametric efficiency bound for the functional $\Psi_{MNAR-A}^a$ under ($\mathcal{M}_{T_1}\cap \mathcal{M}_{T_0})\cap (\mathcal{M}_{\pi_{A,R_A}}\cap \mathcal{M}_{\pi_{R_L}})$.}
  \end{proposition}
We also describe the asymptotic properties of TMLE-A when the nuisance functions are estimated with flexible machine learning algorithms, and consider the challenges in the Discussion. This can be accomplished by estimating the nuisance functions via, e.g., generalized additive models, adaptive regression splines or other flexible machine learning algorithms.

\begin{theorem}[Weak convergence of TMLE-A]
    Suppose that the conditions given in Web Appendix H hold,  and further suppose that the following condition also holds:
    \begin{align*}
     \norm{\hat{T}_1(L) - T_1(L)}_2~&\norm{\hat{\pi}_A(L)\hat{\pi}_{R_A}(L)-{\pi}_A(L){\pi}_{R_A}(L)}_2 + \\
       & \norm{\hat{{T}}_0(L_O) - {T}_0(L_O)}_2~\norm{\hat{\pi}_{R_L}(L_O)-{\pi}_{R_L}(L_O)}_2  = o_p(n^{-1/2}).
    \end{align*}
where $\norm{f(x)}_2 = \left\{\int |f(x)|^2dP(x)\right\}^{1/2}$, i.e. the $L_2(P)$ norm.
Then, $\sqrt{n}(\widehat{\Psi}_{\bf{TMLE,MNAR-A}}^a - \Psi_{MNAR-A}^{a}) \rightsquigarrow N(0,\sigma^2),~~\sigma^2 = \text{Var}\left(\phi^1_{P_{\bf{MNAR-A}}}\right).$ 
\end{theorem}

{As such, when the nuisance functions are estimated with machine learning algorithms, the variance of TMLE-A can be estimated empirically with $\mathbb{P}_n\left[\left(\hat\phi^1_{P_{\bf{MNAR-A}}}\right)^2\right]$, where all nuisance functions are replaced with their estimates. When the nuisance functions are estimated with parametric models, this variance estimator remains applicable provided that all nuisance functions are correctly specified. However, in practice, we recommend using the nonparametric bootstrap to estimate the variance, as we are more susceptible to model misspecification with parametric models.}
By treating missingness in $A$ and $L_M$ separately, we are able to consider a more flexible set of assumptions. In particular, this set of assumptions allows $R_A$ to depend on $A$ and $L_M$. However, it is still quite restrictive with respect to confounder missingness, as $R_L$ still cannot depend on any variables in $L_M$. Again, we can make progress by taking a more granular approach; in the following section, we consider distinct missingness mechanisms for different confounders.

\subsection{A strategy for separating observation indicators in $L_M$}
\label{sec:misssep}

Suppose that there are $q$ variables in $L_M$ such that $L_M=(L_{M1},\ldots,L_{Mq})$. Let $R_{Lk}$ denote the observation indicator for covariate $L_{Mk}$ for $k=1,\ldots,q$ (i.e., $R_{Lk}=1$ if $L_{Mk}$ is observed and $R_{Lk}=0$ otherwise).  
Furthermore, for any random variable $X$ of length $q$, let $\bar{X}_{k} = (X_{1},\ldots, X_{k})$, ${X} \coloneqq \bar{X}_q = (X_{1},\ldots, X_{q})$  and $\underline{X}_{k} = (X_{k},\ldots,X_{q}).$
We now observe $O = (L_O,R_{L1},\ldots,R_{Lq}, R_A, R_{L1}L_{M1}, \ldots,R_{Lq}L_{Mq}, R_A A, Y)\sim P$ and consider the following assumptions for identification in this setting:
\begin{assumption}[MNAR-B]
\label{ass:MNAR-Amon} ~
\begin{enumerate}
    \item {Exposure missingness: $R_A \indep Y|A,L$} 
    \item {Covariate missingness: $(R_{L1},\ldots,R_{L_q}) \indep Y|A,L,R_A$ \text{and} $R_{Lk} \indep \underline{L}_{Mk}\mid \bar{R}_{L,k-1}, \bar{L}_{M,k-1},L_O$}
\end{enumerate}
    $\forall k=1,\ldots,q$
\end{assumption}
\begin{assumption}[Positivity: MNAR-B]
\label{ass:posMNAR-Amon}
         $$P(A=a\mid R=1, L)>0,~~\text{and}~~P(R_{Lk}=1\mid \bar{R}_{L,k-1}=1_{k-1},\bar{L}_{M,k-1}, L_O)>0,~~\forall k=1,\ldots,q$$ 
         where $1_k$ denotes a vector of ones of length $k$, for any $k$.
\end{assumption}

\begin{figure}[!htbp]
	\centering
 \begin{minipage}{1\textwidth}
   \centering
	\scalebox{1}{
	\begin{tikzpicture}
 \node (LO) at (-5,-1) {$L_O$};
		\node (LM1) at (-4,2) {$L_{M1}$};
		\node (LM2) at (-1,2) {$L_{M2}$};
		\node (LM3) at (2,2) {$L_{M3}$};
		\node (RL1) at (-3.5,4) {$R_{L1}$};
        \node (RL2) at (0.5,4) {$R_{L2}$};
        \node (RL3) at (3.5,4) {$R_{L3}$};
		\draw [<->]  (LO) edge[dashed] (LM1);
        \draw [<->]  (LO) edge[dashed] (LM2);
         \draw [<->]  (LO) edge[dashed] (LM3);
		\draw [<->] (LM1) edge [dashed](LM2);
        \draw [<->] (LM2) edge[dashed] (LM3);
        \draw [<->] (LM1) edge[bend right,dashed, out=330, in=195] (LM3);
        \draw [->] (LM1) edge (RL2);
        \draw [->] (LM1) edge (RL3);
        \draw [->] (LM2) edge (RL3);
        \draw [<->]  (RL1) edge[dashed] (RL2);
        \draw [<->]  (RL2) edge[dashed] (RL3);
         \draw [<->]  (RL1) edge[bend left,dashed] (RL3);
         \draw [->]  (LO) edge[bend left=30] (RL1);
         \draw [->]  (LO) edge (RL2);
         \draw [->]  (LO) edge[bend right=40] (RL3);
	\end{tikzpicture} }
  \end{minipage} 
	\caption[.]{Examples of a DAG satisfying missingness assumption $R_{Lk} \indep \underline{L}_{Mj}\mid \bar{R}_{j-1}, \bar{L}_{M,j-1},L_O,$ for $q=3$. Dashed lines represent potential unmeasured common causes.}
  \label{fig:monotone}
\end{figure}
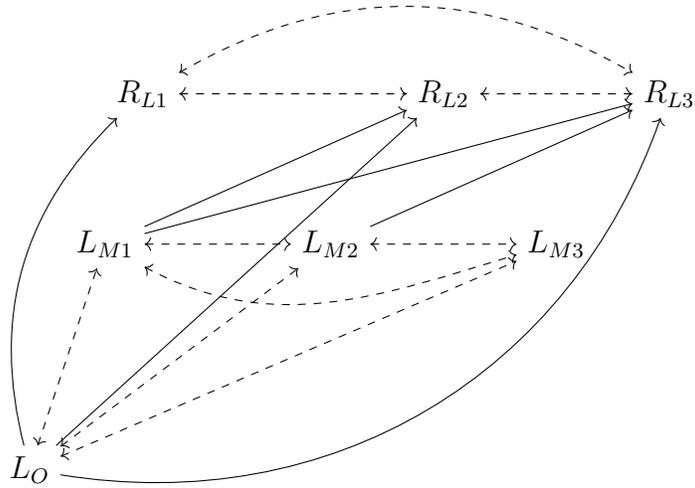

{Note that $R_{Lk} \indep \underline{L}_{Mk}\mid \bar{R}_{L,k-1}, \bar{L}_{M,k-1},L_O$ is equivalent to $\bar{R}_{Lk} \indep \underline{L}_{Mk}\mid \bar{L}_{M,k-1},L_O$, and closely resembles the block-conditional MAR of \cite{zhou2010block} as well as the sequential ignorability assumption of \cite{Bang2005}.}
Figure \ref{fig:monotone} illustrates a scenario satisfying $R_{Lk} \indep \underline{L}_{Mk}\mid \bar{R}_{L,k-1}, \bar{L}_{M,k-1},L_O$ for $q=3$. 
{}
Note that  neither assumption \ref{ass:MNAR-A} nor \ref{ass:MNAR-Amon} imply one another. Specifically, it can be shown that in general, $\prod_{k=1}^q R_{Lk} \indep L_M\mid L_O$ does not imply $R_{Lk} \indep \underline{L}_{Mk}\mid \bar{R}_{L,k-1}, \bar{L}_{M,k-1},L_O,~~\forall k=1,\ldots,q$, nor vice versa. 

Under assumptions \ref{ass:MNAR-Amon} and \ref{ass:posMNAR-Amon}, $E(Y^{a})$ can be identified using the following functional:
\begin{align}
   \Psi_{MNAR-B}^{a} =  \sum_l E(Y\mid A=a, R_A=1, \bar{R}_{Lq}=1, L)\prod_{k=1}^q p(l_{Mk}\mid l_O, \bar{l}_{M,k-1}, \bar{R}_{Lk}=1_k)p(l_O). \label{eq:idMNAR-Amon}
\end{align}
where $\bar{L}_{M0} = \emptyset$ by definition.

The ordering of the confounders plays an important role here. First, the assumptions for $R_{Lj}$ and $R_{Lk}$ differ for $j<k$. In particular, one could order the confounders in $L_M$ to best capture the underlying causal structure where, e.g., $L_{Mj}$ may potentially affect $R_{Lk}$ but $L_{Mk}$ does not affect $R_{Lj}$. In our data application, education was subject to missingness and was thought to potentially affect the missingness status of other covariates like income {(e.g., less educated individuals with limited financial literacy may not know their annual income and thus may leave the income blank on the survey)}. Hence, including education as the first covariate in $L_M$ (i.e., as $L_{M1}$) allows missingness status of income to depend on education. Second, even if data are not monotone missing (monotone in the sense that if $L_{Mk}$ is observed, $\{L_{M1},\ldots,L_{M,k-1}\}$ must be observed, $\forall k$), monotone missingness will be implicitly enforced in estimators that are based on identifying formula (\ref{eq:idMNAR-Amon}).
That is, $L_{Mk}$ will be treated as missing if any of $\{L_{M1},\ldots,L_{M,k-1}\}$ are missing. Hence, in our example, income will be treated as though it were missing whenever education is missing. In the following subsection, we will see how this artificial monotonicity can affect estimation.


\subsubsection{Estimation}
We consider estimation of the functional $\Psi_{MNAR-B}^{a}$ given by identifying formula (\ref{eq:idMNAR-Amon}) using a TMLE motivated by the corresponding influence function for $ \Psi_{MNAR-B}^{a}$ under a nonparametric model $\mathcal{M}_{np}$ for $P$:
\begin{align}
    \phi^1_{P_{\bf{MNAR-B}}} = & \frac{I(A=a,R_A=1,\bar{R}_{Lq}=1_q)}{\pi_A(L)\pi_{R_A}(L)\prod_{k=1}^q \pi_{R_{Lk}}(\bar{L}_{M,k-1},L_O)}\{Y-\tilde{T}_{q}(L)\} +\label{eq:eifmnar12mon} \\ 
    &\sum_{k=1}^{q}\frac{I(\bar{R}_{Lk}=1_k)}{\prod_{j=1}^k \pi_{R_{Lj}}(\bar{L}_{M,j-1},L_O)}\{\tilde{T}_k(\bar{L}_{Mk}, L_O)-\tilde{T}_{k-1}(\bar{L}_{M,k-1},L_O)\} + \tilde{T}_0(L_O) - \Psi_{MNAR-B}^{a}
 \nonumber     
\end{align} 
where the probability of observing each covariate $k$ in $L_M$ is given by:
$$\pi_{R_{Lk}}(\bar{L}_{M,k-1}, L_O)=P(R_{Lk}=1\mid \bar{R}_{L,k-1}=1_{k-1},\bar{L}_{M,k-1}, L_O),$$ for $k=1,\ldots,q$, and the new outcomes are given by:
$$\tilde{T}_k(\bar{L}_{Mk}, L_O) = E\{\tilde{T}_{k+1}(\bar{L}_{M,k+1}, L_O)\mid \bar{L}_{Mk}, L_O,\bar{R}_{Lk}=1_k\}$$ defined iteratively backwards from $k=q-1$ to $k=0$, with $\tilde{T}_{q}(L)=T_1(L)$. 
In Table \ref{alg:tmle_MNAR-Amon}, we present a TMLE motivated by the influence function given by \eqref{eq:eifmnar12mon}.
\vspace{1em}

\begin{table}
\caption{\label{alg:tmle_MNAR-Amon}}
\begin{breakablealgorithm}
\renewcommand{\theenumi}{\Alph{enumi}}   
\caption{{TMLE-B under MNAR-B}}    
\begin{algorithmic} [1]  
\setstretch{1.5}
\item Obtain estimates $\hat\pi_A(L)$ and $\hat\pi_{R_A}(L)$ and $\hat\pi_{R_{Lk}}(\bar{L}_{M,k-1}, L_O)$ of  $\pi_A(L)$, $\pi_{R_A}(L)$ and $\pi_{R_{Lk}}(\bar{L}_{M,k-1}, L_O)$ (for $k=1,\ldots,q$), respectively.

\item \textit{Obtain initial prediction of $\tilde T_q(L)$, $\hat{\tilde T}_q^0(L)$}:

\noindent Among those with $R_A=1$ and $\bar{R}_{Lq}=1_q$, fit a regression model $\zeta(A,L;\kappa)=g^{-1}([A,L]'\kappa)$ by regressing $Y$ on $A$ and $L$. Obtain predictions $\hat{\tilde T}_q^0(L) = \zeta(A,L;\hat\kappa)$ for these individuals.
\item \textit{Targeting step for $\tilde T_q(L)$ to obtain updated predictions $\hat{\tilde T}_q^\ast(L)$:}

\noindent Among those with $A=a$, $R_A=1$ and $\bar{R}_{Lq}=1_q$, regress $Y$ on an intercept with observational weight $\{\hat\pi_A(L)\hat\pi_{R_A}(L)\prod_{k=1}^q\hat\pi_{R_{Lk}}(\bar{L}_{M,k-1}, L_O)\}^{-1}$ and offset given by $g\{\hat{\tilde T}_q^0(L)\}$, i.e., solve for $\epsilon$ in 
\begin{equation*}
   \mathbb{P}_n \left\{ \frac{I(A=a,R_A=1,\bar{R}_{Lq}=1_q)}{\hat{\pi}_A(L)\hat\pi_{R_A}(L)\prod_{k=1}^q\hat\pi_{R_L}(\bar{L}_{M,k-1}, L_O)}\left(Y - g^{-1}\left[g\{\hat{\tilde T}_q^0(L)\}+\epsilon\right]\right)\right\} = 0
\end{equation*}
Among those with $\bar{R}_{Lq}=1_q$, predict $\tilde{T}_q(L)$ using $\hat{\tilde{T}}_q^\ast(L)=g^{-1}\left[g\{\hat{\tilde T}_q^0(L)\}+\hat\epsilon\right]$.
\item Recursively from $k=q,...,1$:
\begin{enumerate}
    \item \textit{Obtain initial prediction of $\tilde T_{k-1}(\bar{L}_{M,k-1},L_O)$, $\hat{\tilde T}_{k-1}^0(\bar{L}_{M,k-1},L_O)$}:
    
    \noindent Among those with $\bar{R}_{Lk}=1_k$, fit a regression model $\eta_{k-1}(\bar{L}_{M,k-1},L_O;\omega_{k-1})=g^{-1}([\bar{L}_{M,k-1},L_O]'\omega_{k-1})$ by regressing $\hat{\tilde{T}}_k(\bar{L}_{Mk},L_O)$ on $\bar{L}_{M,k-1}$ and $L_O$. Obtain predictions $\hat{\tilde T}_{k-1}^0(L) = \eta_{k-1}(\bar{L}_{M,k-1},L_O;\hat\omega_{k-1})$ for these individuals.
    \item  \textit{Targeting step for $\tilde T_{k-1}(\bar{L}_{M,k-1},L_O)$ to obtain updated predictions $\hat{\tilde T}_{k-1}^\ast(\bar{L}_{M,k-1},L_O)$:}
    
    \noindent Among those with $\bar{R}_{Lk}=1_k$, regress $\hat{\tilde{T}}_k^\ast(\bar{L}_{Mk},L_O)$ on an intercept with observational weight $\prod_{j=1}^k\hat\pi_{R_{Lj}}(\bar{L}_{M,j-1}, L_O)^{-1}$ and an offset term given by $g\left\{\hat{\tilde T}_{k-1}^0(\bar{L}_{M,k-1},L_O)\right\}$, i.e., solve for $\nu_{k-1}$ in 
\begin{equation*}
   \mathbb{P}_n \left\{ \frac{I(\bar{R}_{Lk}=1_k)}{\prod_{j=1}^k\hat\pi_{R_{Lj}}(\bar{L}_{M,j-1}, L_O)}\left(\hat{\tilde{T}}_k^\ast(\bar{L}_{Mk},L_O) - g^{-1}\left[g\left\{\hat{\tilde T}_{k-1}^0(\bar{L}_{M,k-1},L_O)\right\}+\nu_{k-1}\right]\right)\right\} = 0
\end{equation*}
Among those with $\bar{R}_{L,k-1}=1_{k-1}$, predict $\tilde{T}_{k-1}(\bar{L}_{M,k-1},L_O)$ using $\hat{ \tilde{T}}_{k-1}^\ast(\bar{L}_{M,k-1},L_O)=g^{-1}\left[g\left\{\hat{\tilde T}_{k-1}^0(\bar{L}_{M,k-1},L_O)\right\}+\hat\nu_{k-1}\right]$.
\end{enumerate}
\item Calculate the TMLE  estimator $\widehat{\Psi}_{\bf{TMLE,MNAR-B}}=\mathbb{P}_n\{\hat{\tilde{T}}_0^\ast(L_O)\}$.
\end{algorithmic}
\end{breakablealgorithm}
\vphantom{}
\end{table}

TMLE-B requires one to specify more models than the estimators proposed earlier, as now we are separately considering the missingness in each incompletely observed covariate. For instance, if we have $q$ incompletely observed covariates, then we will need to specify $q-1$ more missingness models as well as $q-1$ more outcome models compared with TMLE-A. 
{As before, let $\mathcal{M}$ with subscripts 
``${\pi_{A,R_A}}$'' denote models with the correct specification $P(A=a, R_A=1\mid R_L=1, L)$. Let $\mathcal{M}$ with subscripts ``${\tilde T_k}$'' denote models with the correct specification of the nuisance functions defined by $\tilde{T}_k(\bar{L}_{Mk}, L_O)$ (for $k=0,\ldots,q)$, and $\mathcal{M}$ with subscripts 
``${\pi_{R_{Lk}}}$'' denote models with the correct specification of $\pi_{R_{Lk}}(\bar{L}_{M,k-1}, L_O)$ (for $k=1,\ldots,q$).
The proposed TMLE-B is doubly robust in the sense that it is consistent for $\Psi_{MNAR-B}^{a}$ under $(\cap_{k=0}^q \mathcal{M}_{\tilde T_k})\cup (\cap_{k=1}^q \mathcal{M}_{\pi_{R_{Lk}}}\cap \mathcal{M}_{\pi_{A,R_A}})$.}

The asymptotic properties of TMLE-B when the nuisance functions are estimated with machine learning algorithms are given as follows:
\begin{theorem}[Weak convergence of TMLE-B]
    Suppose that the conditions given in Web Appendix H hold, and further suppose that the following condition also holds:
    \begin{align*}
   \sum_{k=1}^q & \norm{\hat{\tilde{T}}_{k-1}(L_O,\bar{L}_{M,k-1}) - {\tilde{T}}_{k-1}(L_O,\bar{L}_{M,k-1})}_2\norm{\hat{\pi}_{R_{Lk}}(L_O,\bar{L}_{M,k-1})-{\pi}_{R_{Lk}}(L_O,\bar{L}_{M,k-1})}_2+ \\
       & \norm{\hat{\tilde{T}}_q(L_O,\bar{L}_{Mq}) - \tilde{T}_q(L_O,\bar{L}_{Mq})}_2~\norm{\hat{\pi}_A(L)\hat{\pi}_{R_A}(L)-{\pi}_A(L){\pi}_{R_A}(L)}_2  = o_p(n^{-1/2}).
    \end{align*}
Then, $$\sqrt{n}\left(\widehat{\Psi}_{\bf{TMLE,MNAR-B}}^a - \Psi_{MNAR-B}^{a}\right) \rightsquigarrow N(0,\sigma^2),~~\text{where}~\sigma^2 = \text{Var}(\phi^1_{P_{\bf{MNAR-B}}}).$$
{The variance of TMLE-B can be estimated as the empirical variance of $\hat\phi^1_{P_{\bf{MNAR-B}}}$ (where all nuisance functions are replaced with their estimates) or via nonparametric bootstrap.}
\end{theorem}

\begin{remark}\label{rmk:morerobustness}
    The proposed TMLE-A and TMLE-B estimators respectively defined under MNAR-A and MNAR-B actually provide more robustness against model misspecification than described herein. See Web Appendix I for a discussion.
 \end{remark}
 \begin{remark}\label{rmk:efficiency1}
  Since  $\Psi_{MNAR-A}^{a}$ and $\Psi_{MNAR-B}^{a}$ are not functions of the exposure or missingness distributions, these TMLEs will attain the efficiency bound under a submodel of the nonparametric model $\mathcal{M}_{np}$ where the observed propensity score and missingness models are known or modelled parametrically \citep{Tsiatis2006}. As such, estimators based on our influence function under correctly specified nuisance functions are guaranteed to be at least as efficient as IPW estimators.
\end{remark}

Clearly, different orderings of variables in $L_M$ correspond to different influence functions for $\Psi_{MNAR-B}^{a}$ identified by \eqref{eq:idMNAR-Amon}.
We have already seen that if the $j^{th}$ covariate in $L_M$ potentially affect the observation status of the $k^{th}$ covariate in $L_M$, then it must be that $j<k$ in order to satisfy assumption \ref{ass:MNAR-Amon}.
However, if there exist a subset of covariates in $L_M$ that do not affect the observation status of any covariate in $L_M$, then there may be different valid orderings, all of which satisfy assumption \ref{ass:MNAR-Amon}. In situations such as these, the ordering of $L_M$ may have important implications for efficiency. 
Without loss of generality, suppose that variables in $L_M$ include `education', `income' and `marital status' (i.e., $q=3$). We let $L_{M1}$ denote `education' as it can potentially affect the observation status of `income' and `marital status', but we hypothesize that `income' and `marital status' do not affect the observation status of any covariate in $L_M$. In this case, the ordering of `income' and `marital status' in $L_M$ is arbitrary, and any ordering will satisfy $(L_{M2},L_{M3})\independent \bar{R}_{L2}\mid L_O,L_{M1}$. However, to make the most use of the available data and potentially optimize efficiency, we may want to place `income' and `marital status' in the order of increasing amount of missingness.

\section{Simulations}\label{s:sims}
We conducted a series of simulations to demonstrate that: (1) MI followed by application of the complete-data TMLE yields valid estimates of the average counterfactual outcome under the MAR assumption, but may yield biased estimates otherwise; (2) the proposed TMLE approaches yield valid inference under the proposed MNAR  assumptions; and (3) the proposed TMLE approaches remain robust against misspecification of either the (a) outcome or (b) exposure and missingness models.

\subsection{Setup and Design}

In each simulated data set, $Y$ was a fully observed binary outcome, $A$ was a partially observed binary exposure, and $L_O$ and $L_M=(L_{M1},L_{M2})$ were common causes of $\{A,Y\}$ that were fully and partially observed, respectively. We allowed for unmeasured common causes of $\{L_O,L_M\}$, $\{L_M,Y\}$ and $\{L_O,Y\}$. 

We generated $1000$ datasets of size $n=2500$ in three scenarios corresponding to different missingness mechanisms. In Scenario I, we observed $\{A,L_M\}$ when $R=1$ (simultaneous missingness). This scenario satisfied the MAR assumption (see Figure \ref{fig:MARwithboth1}[a]; $R$ depended only on $L_O$). We considered separate missingness mechanisms in Scenarios II---III. 
Scenario II satisfied MNAR-A (Figure \ref{fig:assMNAR-A3}[a]; included unmeasured common causes of $\{A,R_A\}$ and $\{A,R_L\}$, and $R_A$ depended on $L_M$ and $A$), and scenario III satisfied MNAR-B (same as II except $R_{L1}$ and $R_{L2}$ are separately generated, and $R_{L2}$ depended on $L_{M1}$).
In each case, the level of missingness was around 20\%--30\%. Full details  are reported in Web Appendix M.

To each dataset we applied five estimators: (i) complete case analysis paired with complete-data TMLE (valid under Missing Completely at Random or MCAR; see Web Appendix J); (ii) MI (via \texttt{mice} package in \texttt{R}) paired with complete-data TMLE estimator (see Web Appendix F); (iii) an iterative conditional expectation (ICE; \citealp{Wen2021}; see also Web Appendix G) estimator for each identifying formulae given throughout; (iv) an IPW estimator for each identifying formulae (see Web Appendix G); and (v) the proposed TMLEs.
{We compared bias, empirical standard error, and  95\% coverage probability (CP; based on nonparametric bootstrap with 1,000 resamples) of each estimator.}

ICE estimators require a model for the outcome process, IPW estimators require models for the exposure and missingness processes, and the proposed TMLEs require all the aforementioned nuisance function models. To investigate robustness to misspecification of these working models, we considered each of these estimators when: (i) all models were correctly specified (i.e., the data-generating models were used), (ii) the outcome model was misspecified, and (iii) the exposure model was misspecified. 
{Details regarding the specification of the parametric models used for the nuisance model can be found in the Web Appendix M.}
Note that we applied TMLE-A in the MAR Scenario I, as MNAR-A was also satisfied.
{For all three scenarios, we also obtained results from TMLE using Highly Adaptive Lasso for fitting the nuisance functions (HAL, \citealp{hejazi2020hal9001}).
Under the condition that the nuisance functions have finite sectional variation norm, HAL can estimate nuisance functional parameters at an approximate rate of $n^{-1/3}$ \citep{van2017generally}.}

\subsection{Simulation Results}\label{ss:simresults}

Table \ref{table:simresultsAll2} shows results for simulation Scenarios I---III. As expected, biases from the complete case estimator are large in all scenarios. {In MAR scenario I, MI is nearly unbiased and the confidence intervals cover roughly at nominal 95\% level. However, in all other scenarios (II-III), MI exhibits greater bias and poor coverage.} In all scenarios, ICE is the most efficient compared with all other estimators.

When all models are correctly specified, IPW, ICE and TMLE are nearly unbiased across all scenarios. Moreover, as long as one set of nuisance function models (outcome models or exposure/missingness models) are correctly specified, TMLE remains nearly unbiased {and achieves close to the nominal 95\% coverage}. In contrast, ICE and IPW exhibit larger biases and lower coverage rates. In Scenarios I---III, TMLEs are at least as efficient as IPW when all the models are correctly specified, as expected from theory (see Remark \ref{rmk:efficiency1}). 
In Web Appendix M, we show an additional simulation study to illustrate efficiency gains discussed in Section \ref{sec:separate} when it is reasonable to place variables in the order of increasing amount of missingness.
{
For Scenarios I--III, the results from TMLE, with nuisance functions estimated using HAL, are given by: (Bias$\times 100$, SE$\times 100$, 95\% CP) = $(0.12,~ 2.51, ~93.7)$, (Bias$\times 100$, SE$\times 100$, 95\% CP) = $(0.06,~ 1.86, ~93.5)$, (Bias$\times 100$, SE$\times 100$, 95\% CP) = $( 0.04,~ 2.09,~93.5)$, respectively, where coverage probability is calculated based on the proposed asymptotic variance estimator.}

\begin{table}[!htbp]
\centering
\begin{tabular}{ |r | rrr | rrr| rrr |  }
\hline
& \multicolumn{3}{|c|}{(i) Correctly } & \multicolumn{3}{|c|}{(ii) Misspecified }  & \multicolumn{3}{|c|}{(iii) Misspecified }  \\
& \multicolumn{3}{|c|}{Specified} & \multicolumn{3}{|c|}{outcome model}  & \multicolumn{3}{|c|}{exposure model}  \\ \hline
\textbf{I (MAR)} \hfill~ 
& Bias  & SE & CP  & Bias  & SE & CP  & Bias  & SE & CP   \\ \hline
CC	&-1.77	&2.84	&85.5	&-1.78	&2.83	&86.0	&-1.77	&2.85	&85.7\\
MI	&-0.04	&2.36	&	95.5 &0.00	&2.36	& 95.4	&-0.04	&2.37	&  95.4\\
ICE-A	&0.09	&2.14	&95.1	&-2.71	&1.89	& 68.1	&0.09	&2.14	&95.1\\
IPW-A	&0.11	&2.56	&94.4	&0.11	&2.56	&94.4	&-4.99	&2.13	& 34.6\\
TMLE-A	&0.12	&2.55	&94.5	&0.11	&2.56	&94.5	&0.11	&2.38	&94.4\\
\hline
\textbf{II (MNAR-A)} \hfill~ & Bias  & SE & CP & Bias  & SE & CP & Bias  & SE & CP  \\ \hline
CC	&1.85	&1.79	&81.8	&1.76	&1.81	&83.0	&1.85	&1.79	&81.8\\
MI	&-1.12	&1.50	& 86.5	&-1.11	&1.52	& 87.7	&-1.12	&1.50	& 86.6\\
ICE-A	&0.04	&1.78	&93.7	&-1.96	&1.69	&75.7	&0.04	&1.78	&93.7\\
IPW-A	&0.01	&1.88	&94.4	&0.01	&1.88	&94.4	&-2.60	&1.80	&65.3\\
TMLE-A	&0.06	&1.87	&94.2	&0.04	&1.87	&94.7	&0.06	&1.84	&94.2\\
\hline
\textbf{III (MNAR-B)} \hfill~ & Bias  & SE & CP & Bias  & SE & CP & Bias  & SE & CP \\ \hline
CC	&3.87	&2.00	&50.1	&3.87	&2.00	&50.1	&3.88	&2.00	&50.0\\
MI	&-1.59	&1.47	& 78.5	&-1.59	&1.47	& 79.3	&-1.59	&1.47	& 78.5\\
ICE-B	&0.00	&1.98	&93.4	&-1.63	&1.91	&83.3	&0.00	&1.98	&93.4\\
IPW-B	&-0.03	&2.11	&94.1	&-0.03	&2.11	&94.1	&-1.40	&2.11	&88.1\\
TMLE-B	&0.03	&2.09	&93.7	&0.01	&2.10	&93.7	&0.03	&2.10	&93.9\\ \hline
\end{tabular}
\caption{Results for simulations I---III for $n=2500$: Bias, standard error (SE), and 95\% confidence interval Coverage Probability (CP) all multiplied by 100. True value of $E(Y^{a=1})=0.288$.}
\label{table:simresultsAll2}
\end{table}

\section{Data Analysis: Opioid Prescriptions and Mortality Among the Elderly}\label{s:analysis}
The first wave of the opioid epidemic in the United States began in the late 1990s, which resulted in a rise of deaths attributed to prescription opioid overdoses \citep{dowell2016cdc}. Recent studies have shown that elderly patients are at a high risk of developing opioid use disorder, partly due to the challenges of pain management, leading to an increased risk of addiction and drug misuse in this population \citep{guerriero2017guidance,dufort2021problematic}. 
Between 2010 and 2015, opioid-related hospitalization increased by 34\% among those 65 years and older \citep{weiss2018opioid}.
Thus, quantifying the effects of opioid prescription on mortality among the elderly is thus an important public health goal.

Using data from the 1999---2004 cycles (i.e., during the first wave of the opioid epidemic) of the National Health and Nutrition Examination Survey (NHANES) study with linkage to mortality databases (National Death Index) through 2015, \citet{inoue2022causal} estimated the effect of opioid prescription on mortality during the 5-year follow-up after the NHANES household interview. As a case study, we investigated the effect of opioid prescription on all-cause death using the same data but restricted to those 65 years and older. 
Our sample contained $n$=3,807 individuals, and included individuals' prescription opioid use ($A\in\{0,1\}$), all-cause mortality ($Y\in\{0,1\}$), and covariates ($L$) including age, sex assigned at birth (male and female), race (non-Hispanic White, non-Hispanic Black, Mexican-American, or others), education levels (less than high school; high school or General Education Degree;  more than high school), poverty-income ratio (the ratio of household income to the poverty threshold), health insurance coverage, marital status, smoking, alcohol intake, and chronic pain status.

As described in \citet{inoue2022causal}, data on prescription medications used in the past 30 days were collected in the in-person interview. Those who responded `yes' to using prescription medication were asked to show their medication containers to the interviewer: if the medication containers were not available, then the interviewer recorded the information verbally reported by the interviewee. 
Opioids identified through this process include codeine, fentanyl, oxycodone, pentazocine and morphine. About 6.5\% of the observed individuals in the sample reported using prescription opioids, and approximately 6\% of the individuals did not show their prescription medication container to the interviewer. Detailed data description can be found in \citet{inoue2022causal}. 

Complete outcome data were available, but data were partially missing for prescription opioid use ($n=43$), education ($n=19$), marital status ($n=117$), poverty-income ratio ($n=410$), smoking status ($n=10$), alcohol intake ($n=257$), health insurance coverage ($n=46$), chronic pain status ($n=16$), with 24\% of the subjects missing either the exposure or a confounder.
We estimated the causal contrast $E(Y) - E(Y^{a=0})$, which quantifies how the observed risk of death differs from the risk of death had no one taken prescription opioids. 
Since some participants might not have reported prescription opioids use despite taking them (possibly illicitly), we further conducted a sensitivity analysis in which subjects who did not show their prescription medication container to the interviewer were treated as having missing exposures (for a total of $n=242$ missing exposure measurements). 

{Whereas \citet{inoue2022causal} handled missing data via MI, we compared: complete case analysis and MI (both followed by the usual complete-data TMLE), as well as our proposed TMLE estimators, as part of a sensitivity analysis. 
We conjecture that prescription opioid users may be less likely to report their true exposure status due to the nature of the in-person interview, hence the MAR assumption---which assumes $A$ cannot affect $R_A$---may be unrealistic here.
To address this, we applied TMLE-A under the MNAR-A assumption \ref{ass:MNAR-A}, treating all of $L_M$ as unobserved when at least one of the covariates is unobserved.
 We further hypothesize that, in addition to demographic variables such as ``age'', ``sex'' and ``race'' which are fully observed, ``education" may also influence the observation of other potential confounders in $L_M$. Thus, we also applied TMLE-B under the MNAR-B assumption \ref{ass:MNAR-Amon} by including ``education'' as the first element of $L_M$ to accommodate this conjecture.   TMLE-B
 implicitly enforces a monotonic missingness in the covariates $L_M$: here education is included first due to concerns outlined above. Absent any concerns about further confounders affecting missingness, the remaining confounders are included in the order of missingness.}  
All 95\% confidence intervals (CIs) were based on the 2.5th and 97.5th percentiles of a nonparametric bootstrap procedure with 5,000 resamples.

We report results for the main and sensitivity analysis in Table \ref{table:analysis}. The MI estimate was very close to the complete case estimate (0.24\% vs 0.23\%) and the 95\% CI of both estimators covered zero. 
That said, the MAR assumption may be violated here due to the sensitive nature of the exposure, and the TMLE estimates (under MNAR-A and MNAR-B) were consistently twice as large, albeit with 95\% CIs that still covered zero. Even the largest estimate, that of TMLE-A, was small (0.62\%, 95\% CI: [$-$0.12,1.39]), which is unsurprising given that we are comparing the risk of death had no one taken prescription opioids to the \textit{observed} risk---and the observed prevalence of prescription opioid use was only 6.5\%. The sensitivity analyses showed a similar pattern of results but with larger effect estimates overall, with the TMLE effect estimate under MNAR-A as high as 0.98\% (95\% CI: 0.10\%, 1.83\%) and under MNAR-B as high as 0.92\% (95\% CI: 0.05\%,  1.74\%).
{In Web Appendix N, we provide results from TMLE when nuisance functions are estimated using HAL, which closely mirror those obtained here with parametric models.}

\begin{table}[!htbp]
\centering
\begin{tabular}{|r | cccc |  }
\hline
&\multicolumn{4}{c|}{\textbf{Main analysis}}\\
& CC & MI  & 
TMLE$_{\bf{MNAR-A}}$ & TMLE$_{\bf{MNAR-B}}$\\ \hline
Estimate & 0.23\% & 0.21\% & 0.62\% & 0.55\%\\
95\% CI & $(-0.19,  0.66)$ & $(-0.17,  0.65)$  & $(-0.12,  1.39)$ & $(-0.19, 1.31)$\\ \hline
& \multicolumn{4}{c|}{\textbf{Sensitivity analysis}}\\
& CC & MI  & 
TMLE$_{\bf{MNAR-A}}$ & TMLE$_{\bf{MNAR-B}}$\\ \hline
Estimate & 0.25\% & 0.20\% &  0.98\% & 0.92\%\\
95\% CI & $(-0.20,0.69)$ & $(-0.17,  0.71 )$ &  $(0.10,  1.83)$ & $(0.10,  1.77)$\\
\hline
\end{tabular}
\caption{Results for data analysis using the NHANES study (three cycles from 1999---2004). CC denotes complete case analysis; TMLE$_{\bf{MNAR-A}}$ and TMLE$_{\bf{MNAR-B}}$  denote TMLE estimators for the identifying formulae under MNAR-A and MNAR-B, respectively. All results are multiplied by 100 (in \%). {All 95\% confidence intervals (CIs) were based on the 2.5th and 97.5th percentiles of a nonparametric bootstrap procedure with 5,000 resamples.}}
\label{table:analysis}
\end{table}

\section{Discussion}\label{s:discussion}
Our work is motivated by missingness in both exposures and confounders, but it is nevertheless instructive to compare with methods for handling missinginess in one of these alone. When confounders alone are missing---and unless they can safely be assumed MAR---identifying average counterfactual outcomes requires further assumptions. Some methods have relied on existence of a shadow variable \citep{miao2018identification} or on the assumption that missing covariates act as confounders only when they are observed \citep{blake2020estimating,blake2020propensity}. More commonly, methods have assumed that missingness is independent of the outcome \citep{ding2014identifiability}---however, this assumption alone is not sufficient for identification. Previous methods relying on this assumption further require a rank constraint---one that precludes the use of binary outcomes, and sometimes discrete outcomes altogether when there are multiple partially observed confounders \citep{ding2014identifiability,yang2019causal,guan2019unified,sun2021semiparametric}. {We too make an outcome-independence assumption, but drop the rank constraint in favour of a further assumption about the confounder missingness mechanism, which, as a consequence, allows our estimators to be applied to outcome types of any kind. The cost of this trade-off is that we make stricter assumptions about missingness in the confounders. 
Future work will explore ways to integrate the no self-censoring models \citep{shpitser2016consistent,nabi2020full,malinsky2022semiparametric} into the missingness assumptions for the $L_M$ variables.}

  {Note that the outcome-independence assumption (and therefore the MNAR assumptions proposed herein) may impose restrictions on the observed-data distribution (e.g., when outcome is continuous but covariates are discrete; \citealp{bartlett2014improving}).
  However, we share the same sentiment as \cite{bartlett2014improving}, who write ``we tend not to worry about [the potential testability of the outcome-independence assumption] in more realistic settings where power to refute the assumption will typically be very low''.
  Future work will explore methods for testing the proposed missing data models in a similar vein as \cite{mohan2014testability} and 
 \cite{nabi2023testability}.
 As such, it may be possible to find semiparametric efficient influence functions by projecting the nonparametric efficient influence functions derived herein onto the tangent space of semiparametric models ($\mathcal{M}_{semi\ast}$), which incorporates any restrictions on the observed data law imposed by the different sets of MNAR assumptions.
 Estimators based on these semiparametric efficient influence functions will attain semiparametric efficiency bounds under $\mathcal{M}_{semi\ast}$, which may be smaller than the nonparametric efficiency bounds under $\mathcal{M}_{np}$. We do not pursue this further in this paper and defer it as an open problem for future research.}

The missingness assumptions and methods considered herein can also be adapted to data with missing values in all variables including the outcome (see Web Appendix L).
Moreover, in more specific scenarios when the exposure alone is missing, previous work relied on the MAR assumption; we are unaware of methods for  estimating marginal average causal effects when the exposure is MNAR. 
In Web Appendix K, we examine the special case where only the exposure is subject to missingness and propose a TMLE estimator under the outcome-independence assumption.

When missingness is disjoint---i.e., some subjects are missing only exposure or only confounders---it is natural to treat $R_A$ and $R_L$ as distinct. But when $R_A$ and $R_L$ are highly correlated---i.e., most subjects are missing both or neither of $A$ and $L_M$---analysts might  opt to handle them as simultaneously missingness as in Section \ref{s:simultaneous} (by treating both $A$ and $L_M$ as missing whenever $R_A \times R_L=0$). The advantages of this approach are that (a) one needs not worry about how observation indicators are related with one another, and (b) it may be easier to reason about the missingness mechanisms and requisite assumptions (i.e., via simpler DAGs). 
By contrast, treating missingness in the exposure and confounders separately as in Section \ref{sec:separate} has the key advantage of loosening restrictions about the missingness mechanisms of the exposure variable.
An analogous decision must be made about whether to treat missingness in different confounders separately. Treating confounders as either all missing or all observed as in Section \ref{sec:MNARA} requires one to specify fewer missingness models, but treating them separately as in Section \ref{sec:misssep} allows for some partially missing confounders to affect whether other confounders are observed. In particular, analysts may order the elements of $L_M$ to best leverage this additional flexibility. 
In general one need not choose between either extreme---lumping all incomplete data together, or treating each incomplete variable separately. One might instead opt for a balance of the two, for example treating education separately because it may affect missingness in other covariates, while treating all other covariates as entirely observed or missing simultaneously.
{In practice, we suggest that it may be sensible to apply the various TMLE methods described herein to a data set as a simple form of sensitivity analysis. This can help determine whether the overall findings are robust to different methodological approaches.}

{Our simulation studies show that all proposed estimators yield valid inference, and ICE estimators are most efficient, when models are correctly specified. We also find that when models may be misspecified, the proposed TMLEs provide more than one chance to achieve valid inference. In practice, nuisance models are generally unknown, and machine learning algorithms can be used to estimate the nuisance functions in any of the TMLE estimators. 
    These estimators can achieve $\sqrt{n}$-consistency and asymptotic normality provided that the estimators of nuisance functions are consistently estimated at a rate faster than $n^{-1/4}$ \citep{Robins2008,Robins2016,Chernozhukov2018}, and our simulations indicated good performance in finite samples. We note, however, that consistency is not guaranteed when more flexible algorithms such as neural network or random forest are used. In these cases---when standard Donsker conditions are not met---sample splitting and cross-fitting can be used to satisfy necessary empirical process conditions \citep{Van2000,Chernozhukov2018}.} 

{In this work, we rely on the conditional exchangeability assumption as a sufficient condition for identifying our causal estimand $E(Y^a)$. However, weaker or alternative assumptions, such as those in proximal learning \citep{tchetgen2020introduction}, may be applicable depending on the setting and available information. Addressing missing data under such frameworks is an important direction for future research.}

\section*{Acknowledgement}
\noindent The authors would also like to thank Dr. Shaun Seaman for insightful comments on this manuscript.
Lan Wen is supported by the Natural Sciences and Engineering Research Council of Canada (NSERC) Discovery Grant [RGPIN-2023-03641, DGECR-2023-00455]. Glen McGee is supported by the Natural Sciences and Engineering Research Council of Canada (NSERC) Discovery Grant [RGPIN-2022-03068, DGECR-2022-00433].

\bibliographystyle{biorefs}
\bibliography{bibliography}

\end{document}